\documentclass[reprint,NumberedRefs]{JASA-EL}

\usepackage{amsmath,amssymb,amsfonts}
\usepackage{graphicx}
\usepackage{textcomp}
\usepackage{xcolor}
\usepackage{xfrac}
\usepackage{enumitem}
\usepackage{bm}
\usepackage{url}
\usepackage{notation}
\usepackage{mathtools}
\usepackage[normalem]{ulem}
\usepackage{multirow}

\newcommand{\ist}{\hspace*{.3mm}}
\newcommand{\rmv}{\hspace*{-.3mm}}

\newcommand{\stkout}[1]{\ifmmode\text{\sout{\ensuremath{#1}}}\else\sout{#1}\fi}

\newcommand{\zs}{z_{\text{s}}}
\newcommand{\zr}{z_{\text{r}}}
\begin{document}

\title{Bayesian single-hydrophone geoacoustic inversion in shallow water using ship noise}

\author{Junsu Jang}
\email{junsu.jang@whoi.edu}
\affiliation{Woods Hole Oceanographic Institution, Woods Hole, MA, 02543, USA}
\author{Julien Bonnel}
\email{jubonnel@ucsd.edu}
\affiliation{Scripps Institution of Oceanography, University of California San Diego, La Jolla, CA, 92093, USA}
\author{Florian Meyer}
\email{flmeyer@ucsd.edu}
\affiliation{Scripps Institution of Oceanography, University of California San Diego, La Jolla, CA, 92093, USA}
% \date{\today} 

\begin{abstract}
We present a Bayesian framework for geoacoustic inversion using relative dispersion curves extracted from modal interference generated by continuous ship noise recorded on a single stationary hydrophone. The Bayesian formulation enables posterior uncertainty quantification and model-order selection using the Bayesian information criterion. Applied to data from the Seabed Characterization Experiment 2017 (SBCEX17), the inversion identifies two finite sediment layers over a half-space and recovers geoacoustic properties consistent with previous studies, establishing relative dispersion curves as an informative observable for passive Bayesian geoacoustic inversion with a single hydrophone.
\end{abstract}

\maketitle

%%%%% SECTION I : Introduction %%%%%
\section{Introduction}
Accurate knowledge of seabed geoacoustic properties is essential for underwater source localization and acoustic signal interpretation in both maritime and ecological applications. Geoacoustic inversion provides an effective means of estimating these properties \cite{ChaSha:J21,MeyHodSau:J26}. Traditional approaches based on active acoustic sources, such as matched-field inversion, provide accurate environmental estimates but require dedicated acoustic transmissions \cite{DosWilKnoBon:J25a,ParKroMeyGer:J25}. Passive approaches are therefore attractive. In particular, hydrophone arrays provide additional spatial information and have enabled successful passive geoacoustic inversion \cite{TolDosKno:J19}, but their deployment cost and logistical complexity can be prohibitive. This has motivated continued interest in geoacoustic inversion methods using a single hydrophone.

Signal-processing methods that exploit the modal structure of shallow-water propagation have proven particularly effective for single-hydrophone geoacoustic inversion. For impulsive (broadband) sources, modal arrivals can be separated using time-frequency methods such as warping, and their dispersion can then be inverted for seabed properties \cite{BonCha:J11,BonDosEleCha:J20}. Continuous passive sources, however, do not produce isolated modal arrivals and thus require alternative observables. Ship-radiated noise provides sustained excitation over a broad frequency band, generating observable modal interference over long propagation ranges\cite{JanHodMey:J25}. Previous studies have exploited ship noise for passive geoacoustic inversion using hydrophone arrays through cross spectra\cite{KocKno:J05} and Bayesian trans-dimensional inversion\cite{TolDosKno:J19,TolHodDosBonKno:J22}. Single-hydrophone approaches have also been proposed, including methods based on low-frequency singular-point structures \cite{KnoWilNeiHod:J21,KnoNeiWilHodBonLin:J22} and optimization-based inversion of relative dispersion curves (RDCs) extracted from modal interference \cite{GerKinBonSteVal:J12}. 

In this letter, we make two contributions. First, we formulate RDC-based geoacoustic inversion within a Bayesian framework. While previous RDC-based inversion was optimization-based and estimated a single bottom sound speed\cite{GerKinBonSteVal:J12}, the Bayesian formulation provides posterior uncertainty quantification for the parameters and enables selection of the number of sediment layers using the standard Bayesian information criterion (BIC)\cite{StoSel:J04}. Second, we demonstrate the approach using ship noise recorded on a single hydrophone of vertical line array 2 (VLA2) during the Seabed Characterization Experiment 2017 (SBCEX17; Fig. 1)\cite{WilKnoNei:J20}.

\begin{figure}[!b]
    \centering
    \begin{minipage}[c]{0.5\linewidth}
        \centering
        \includegraphics[width=\linewidth]{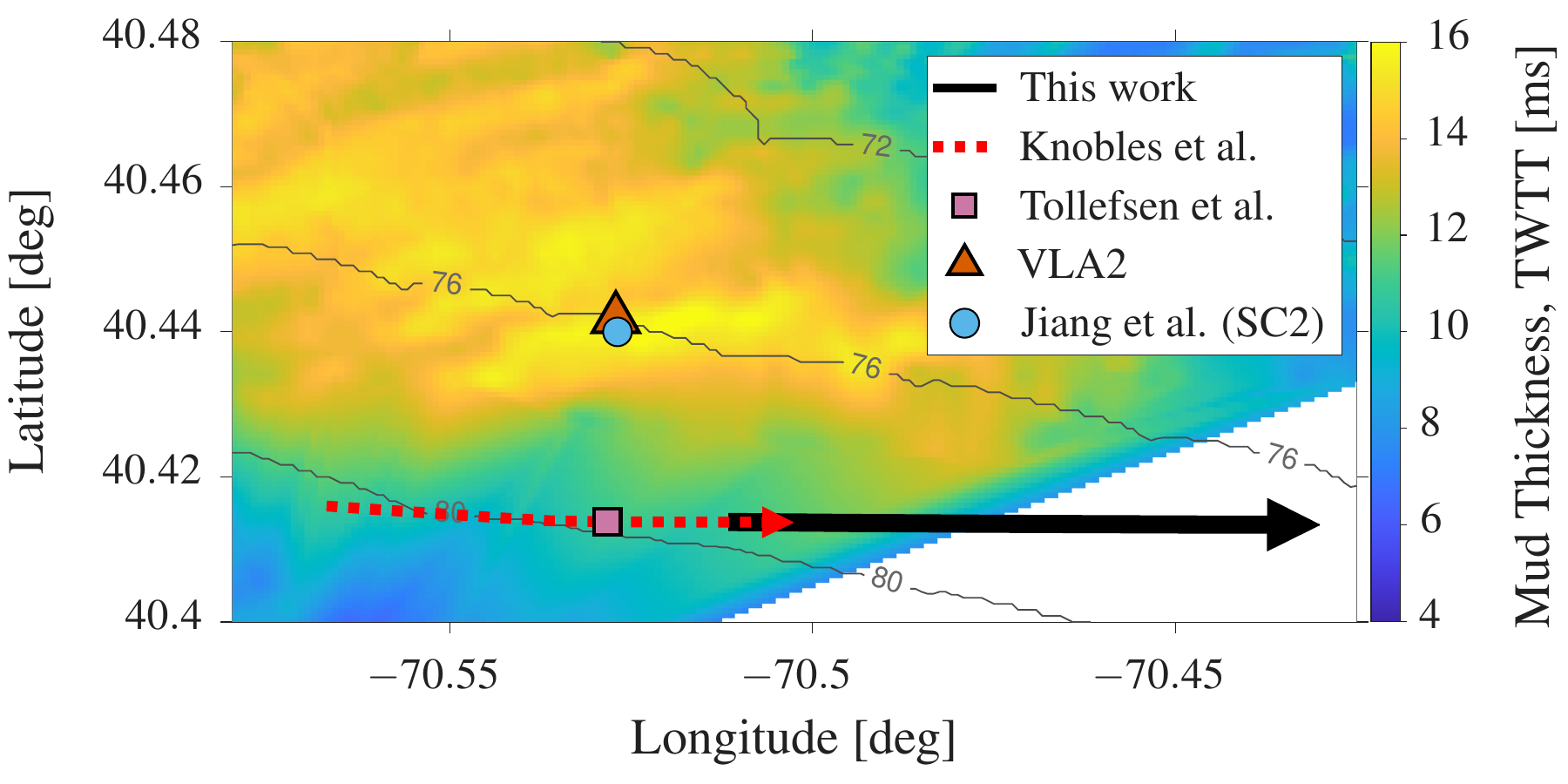}
    \end{minipage}\hfill
    \caption{MSC KALAMATA track and locations associated with the geoacoustic models considered in this study. The black and red track segments indicate the portions of the MSC KALAMATA pass recorded on VLA2 used in this work and by Knobles et al.\cite{KnoNeiWilHodBonLin:J22}, respectively. The square indicates the ship location at closest point of approach (CPA) for the recording used by Tollefsen et al.\cite{TolHodDosBonKno:J22}. The VLA2 location is marked by the triangle, while the SC2 location (circle) marks the local geoacoustic profile of Jiang et al. \cite{JiaHolDosDet:J23}. Bathymetric contours (thin black lines) are shown in meters, while the color background indicates mud thickness in two-way travel time (TWTT) \cite{GofReeGawWilKno:J19}.}
    \label{fig:scenario}
\end{figure}

\section{Methods}
% Background on modal interference and RDCs
\subsection{Modal Interference and Relative Dispersion Curves}
\begin{figure}[htbp]
    \centering
    \includegraphics[width=0.32\linewidth]{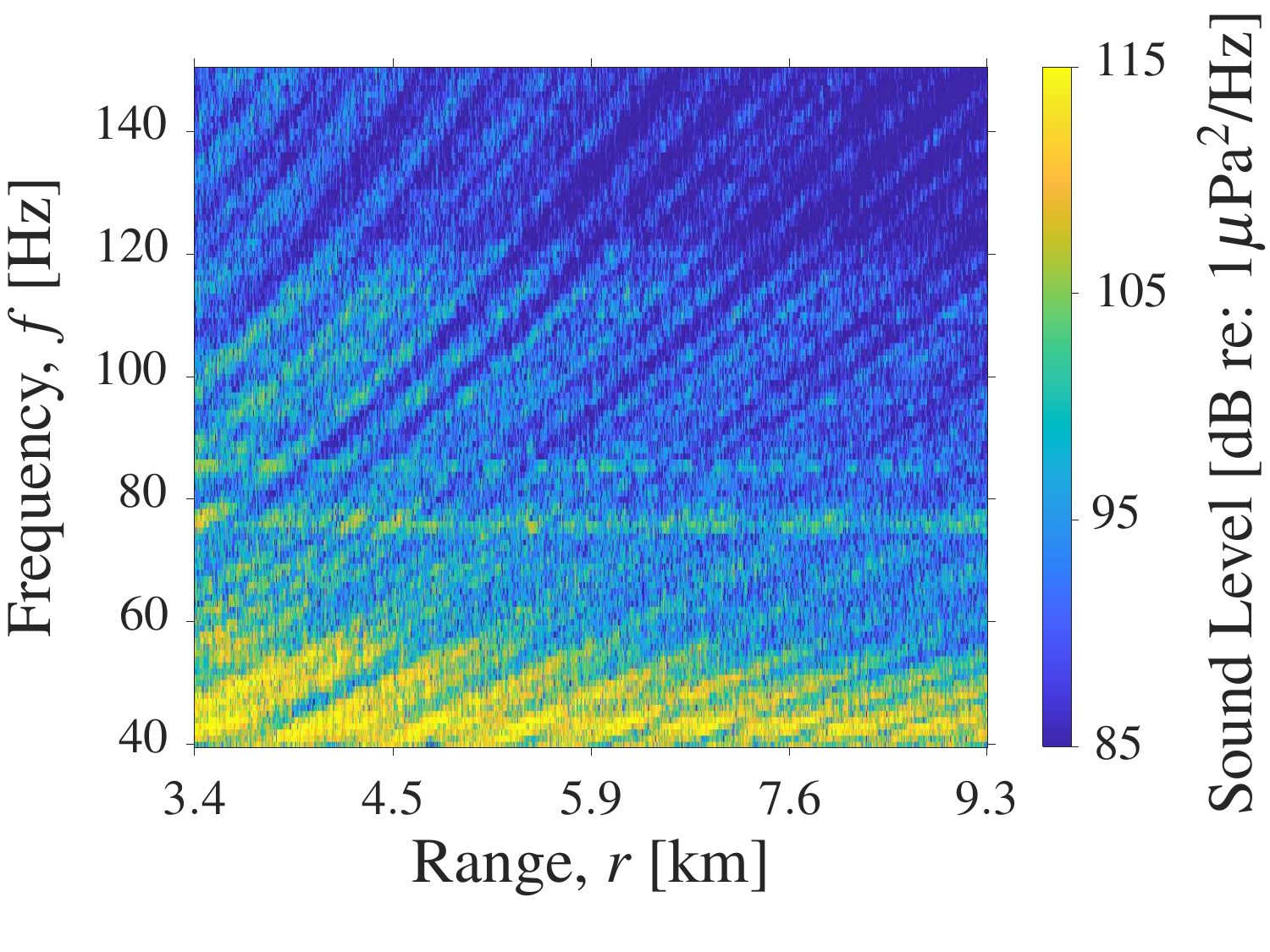}\hfill
    \includegraphics[width=0.32\linewidth]{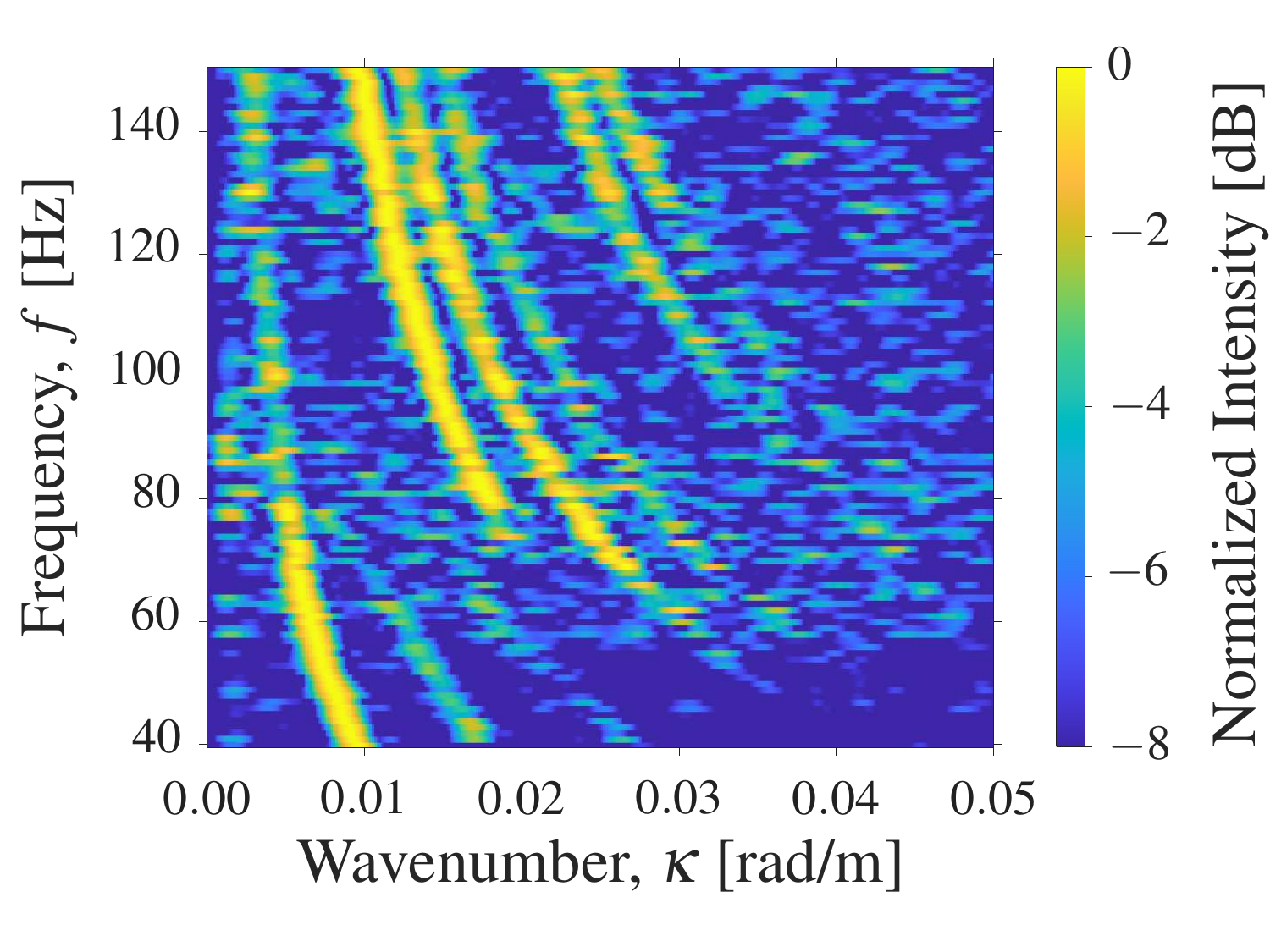}\hfill
    \includegraphics[width=0.32\linewidth]{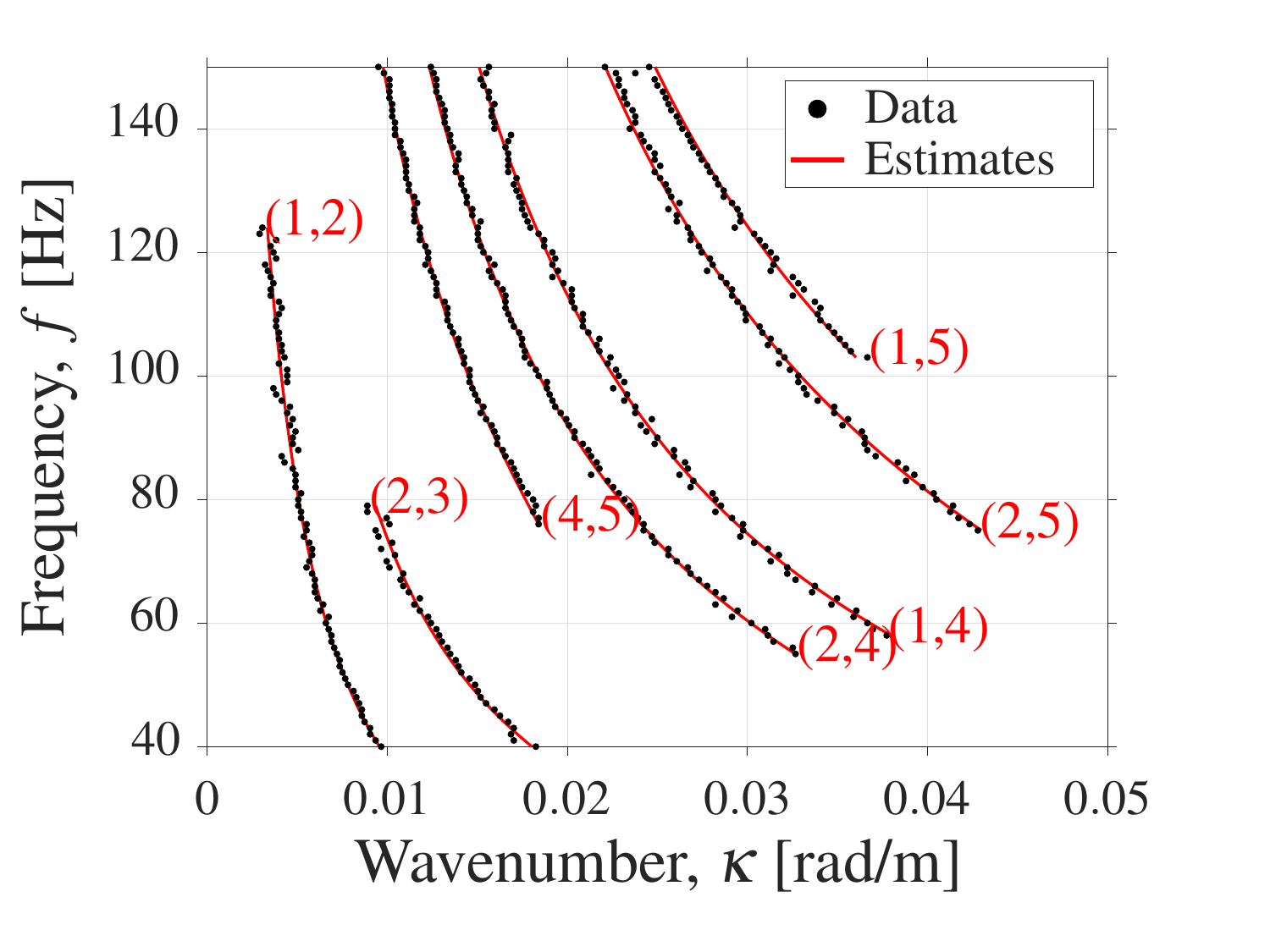}
    \caption{(a) Ship-noise spectrogram recorded during the passage of the container ship MSC KALAMATA during SBCEX17 using a single hydrophone of VLA2. (b) Wavenumber representation obtained by Fourier transforming the range-domain spectrogram, showing the relative dispersion curves (RDCs). The plot is normalized at each frequency. (c) Extracted RDC measurements (black dots) together with the predicted RDCs corresponding to the posterior-mean parameter vector (red curves). The labels indicate the associated mode pairs.}
    % \vspace{-.4in}
    \label{fig:data_ex}
\end{figure}

In a range-independent waveguide, the frequency-domain pressure can be represented as a sum of normal modes\cite{JenKupPorSch:B11_WI}. Let $A_m=A_m(f,\zs,\zr)$ denote the generally complex amplitude of mode $m$ at frequency $f$, source depth $\zs$, and receiver depth $\zr$, and let $k_m=k_m(f)$ denote its real horizontal wavenumber, neglecting modal attenuation. The acoustic intensity at range $r$ can then be written approximately as
\begin{equation}
    I(r,f) \propto \frac{P(f)}{r}\Biggl[\sum_m \lvert A_m\rvert^2
    +2\sum_{m<n}\operatorname{Re}\!\left\{A_m A_n^*
    \exp\!\left(\mathrm{i}[k_m-k_n]r\right)\right\}\Biggr].
    \label{eq:mode_eq}
\end{equation}
Here, $P(f)$ is the source power-spectrum term, and ${}^*$ denotes complex conjugation. Each modal cross term oscillates with angular spatial wavenumber $\Delta k_{mn}(f)=\lvert k_m(f)-k_n(f)\rvert$ (Fig.~\ref{fig:data_ex}(a)). After compensating for cylindrical spreading by multiplying by $r$, the range transform over the interval [$r_{\min},r_{\max}$] is defined as\cite{GerKinBonSteVal:J12}
\begin{equation}
    \widetilde I(\kappa,f)
    =\int_{r_{\min}}^{r_{\max}} r I(r,f)
    \exp\!\left(-\mathrm{i}\kappa r\right)\,\mathrm{d}r.
    \label{eq:continuous_range_transform}
\end{equation}
In practice, the measured intensity was first resampled onto a uniformly spaced range grid, and this transform was evaluated using a discrete Fourier transform. The finite range aperture $5.9$~km yields a wavenumber resolution of $1.06\times10^{-3}$~ rad/m. To reduce quantization in peak picking, the spectrum was zero-padded to a grid spacing of $0.15\times10^{-3}$~rad/m (Fig.~\ref{fig:data_ex}(b)). The resulting spectrum contains peaks near $\kappa=\Delta k_{mn}(f)$, and since $r I(r,f)$ is real, only the nonnegative spatial-wavenumber spectrum is shown in Fig.~\ref{fig:data_ex}(b). Each sufficiently resolved high-intensity ridge defines an RDC associated with a mode pair $(m,n)$, where $m<n$. The geoacoustic inversion compares RDCs predicted by a normal-mode model with labeled RDCs extracted from ship-noise data (Fig.~\ref{fig:data_ex}(c); mode-pair labeling is described in Sec.~\ref{sec:geo_inv}).

% Fig. 2
% spectrogram --> RDC in one of the hydrophones

\subsection{Geoacoustic Inversion}\label{sec:geo_inv}
We employ a Bayesian framework to estimate an unknown but fixed $d$-dimensional geoacoustic parameter vector $\V{\theta}\in\mathbb{R}^{d}$. Let $\mathcal{M}\subset\{(m,n)\in\mathbb{N}^2:m<n\}$ denote the set of observed mode pairs, and let $\mathcal{F}_{mn}=\{f_{mn}^{(1)},\ldots,f_{mn}^{(N_{mn})}\}$ denote the frequencies over which the RDC associated with mode pair $(m,n)\in\mathcal{M}$ is observed. The mode-pair identities are not included in $\V{\theta}$. Instead, they were assigned manually prior to inversion by matching the ordering of the observed RDCs in $\kappa$ across frequency with that of the RDCs predicted using the reference geoacoustic model of Ref.~\citen{JiaHolDosDet:J23}. Although the predicted and measured wavenumber differences did not coincide numerically, their ordering was assumed to be preserved. The total number of measurements is $N=\sum_{(m,n)\in\mathcal{M}}N_{mn}$. The observation corresponding to mode pair $(m,n)$ at frequency $f_{mn}^{(i)}$, where $i=1,...,N_{mn}$, is the measured modal wavenumber difference, $y_{mn}^{(i)}=\Delta k_{mn}^{\mathrm{obs}}(f_{mn}^{(i)}).$

For a trial geoacoustic parameter vector $\V{\theta}$, the modal horizontal wavenumbers are computed using KRAKENC, a normal-mode program that solves for complex modal wavenumbers in attenuating waveguides~\cite{Por:R92}; their real parts are used to form the predicted RDCs. The forward model predicts the RDC as the modal-wavenumber difference, $g_{mn}(f_{mn}^{(i)};\V{\theta})
    =
    \left|
    k_m(f_{mn}^{(i)};\V{\theta})
    -
    k_n(f_{mn}^{(i)};\V{\theta})
    \right|$.
The measurements are modeled as $
    y_{mn}^{(i)}
    =
    g_{mn}(f_{mn}^{(i)};\V{\theta})
    +
    \eta_{mn}^{(i)}$,
where $\eta_{mn}^{(i)}\sim\mathcal{N}(0,\sigma_{mn}^2).$
A single standard deviation \(\sigma_{mn}\) is used across all frequencies for each mode pair, with errors assumed independent across frequencies and mode pairs.

The posterior distribution is obtained from Bayes' theorem,
    $p(\V{\theta}\mid\V{y})
    \propto
    p(\V{y}\mid\V{\theta})\ist
    p(\V{\theta})$,
where $\V{y}\in\mathbb{R}^{N}$ denotes the complete measurement vector. We assign mutually independent uniform priors, $\theta_j\sim\mathcal{U}(a_j,b_j)$ for $j=1,\ldots,d$, where $[a_j,b_j]$ is the bound for component $\theta_j$. Under the measurement model above with additive Gaussian error, the likelihood is
\begin{equation}
    p(\V{y}\mid\V{\theta})
    =
    \prod_{(m,n)\in\mathcal{M}}
    (2\pi\sigma_{mn}^2)^{-N_{mn}/2}
    \exp\!\left[
    -\frac{1}{2\sigma_{mn}^2}
    \sum_{i=1}^{N_{mn}}
    \left(
    y_{mn}^{(i)}
    -
    g_{mn}(f_{mn}^{(i)};\V{\theta})
    \right)^2
    \right].
\end{equation}
Although the measurement errors are modeled as Gaussian, the posterior distribution is generally non-Gaussian because the forward model is nonlinear in $\V{\theta}$ and the uniform priors truncate the posterior at the prior bounds.

We estimate the geoacoustic parameter $\theta$ using Markov chain Monte Carlo (MCMC) \cite{RobCas:05}. We employ an adaptive Metropolis--Hastings algorithm\cite{HaaSakTam:J01} with covariance adaptation and principal-component sampling to improve convergence and sampling efficiency in correlated and multimodal posterior distributions \cite{MeyHodSau:J26}. We summarize the inversion using the minimum mean-square error (MMSE) estimate, i.e., $\widehat{\V{\theta}}_{\mathrm{MMSE}}=\mathbb{E}[\V{\theta}\mid\V{y}]=\int_{\Theta}\V{\theta}\,p(\V{\theta}\mid\V{y})\,\mathrm{d}\V{\theta}$, together with the marginal 95\% credible intervals (CIs) of each parameter.

\subsection{Model Selection}
To determine the appropriate sediment parameterization, we compare models with different numbers of sediment layers using the BIC \cite{StoSel:J04}. In this work, the model order $q$ is the number of finite sediment layers over a half-space; each additional finite layer introduces three unknown parameters (thickness, sound speed, and density). The BIC balances model fit against model complexity and is defined as $\mathrm{BIC}_q
    =
    -2\ln p(\V{y}\mid\hat{\V{\theta}}_{q,\mathrm{ML}},\mathcal{H}_q)
    +
    d_q\ln N$,
where $\hat{\V{\theta}}_{q,\mathrm{ML}}$ denotes the maximum-likelihood estimate under the candidate model $\mathcal{H}_q$, $d_q$ is the number of independently estimated parameters, and $N$ is the number of RDC observations; smaller values are preferred. The BIC approximates $-2\ln p(\V{y} \mid \mathcal{H}_q)$, where the evidence $p(\V{y} \mid \mathcal{H}_q)$ is the likelihood averaged over the prior. Each additional parameter multiplies the evidence by approximately the ratio of its posterior width to its prior width; because this ratio scales as $1/\sqrt{N}$, it yields the $\ln N$ penalty.

\subsection{Dataset and Implementation Details}\label{sec:impl}
We process data collected during SBCEX17 near the New England Mud Patch (NEMP). The acoustic data were recorded on a stationary 16-element VLA2 deployed by the \textit{Marine Physical Laboratory} at a nominal water depth of 76~m. We consider the 13\textsuperscript{th} hydrophone, which was the deepest element providing usable data and located at a nominal depth of 60~m.

The analyzed recording spans a 12-min interval from 18:33 to 18:45 UTC on 24 March 2017, during which the container ship MSC KALAMATA (Maritime Mobile Service Identity: 477510600) traveled away from the closest point of approach (CPA) relative to VLA2 (Fig.~\ref{fig:scenario}). The post-CPA segment provided an approximately monotonic range evolution. During this interval, the ship maintained an approximately constant speed of 20~kn and the source--receiver range increased from $r_{\min} = 3.4$ to $r_{\max}=9.3$~km. The propagation environment was mildly range dependent due to bathymetric variation (Fig.~\ref{fig:scenario}), with the water depth beneath the ship decreasing from 79~m at the CPA to 77~m at the end of the analyzed segment.

The spectrogram was computed using 1~s Hann windows and interpolated from time to range using ship positions from the Automatic Identification System. The range-domain signal was sampled every 2.5~m. RDCs were extracted using the Radon-transform approach of Ref.~\citen{GalBon:J13b}: at each frequency, the RDC measurement was taken as the peak within $\pm0.0005$~rad/m of the ridge identified in the $(f,\kappa)$ plane (Fig.~\ref{fig:data_ex}(c)). The measurement-error standard deviations \(\sigma_{mn}\) were estimated from the deviations of these peaks from the ridges, yielding $(2.1, 3.4, 1.9, 2.6, 3.0, 2.7, 4.1)\times 10^{-4}$~rad/m for mode pairs (1,2), (2,3), (4,5), (2,4), (1,4), (2,5), and (1,5). Only RDCs between 40 and 150~Hz were used.

Geoacoustic inversion was performed using models with $q=1$, $q=2$, and $q=3$ finite sediment layers over a half-space. For a model of order $q$, the parameter vector $\V{\theta}$ comprised the water depth, thicknesses of the $q$ finite sediment layers, and the compressional sound speeds and densities of the sediment layers and half-space. The prior bounds for the preferred q = 2 model are summarized in Table~\ref{tab:results}; the q = 1 model used the corresponding subset of parameters. The sound-speed bounds for the surficial layer, [1400, 1600] m/s, include values below the water sound speed, allowing for the low-speed mud known to be present at the NEMP. For the q = 3 model, the third finite layer, located beneath the second layer and above the half-space, was assigned uniform prior bounds of [2.0, 100.0] m, [1500, 2200] m/s, and [1.4, 2.6] g/cm³ for thickness, sound speed, and density, respectively. The water column was assumed isovelocity at $1472.5$~m/s based on field measurements, and the attenuation coefficients of the sediment layers and half-space were fixed at $0.05$~dB/(m$\cdot$kHz). After discarding burn-in samples, the $q=1$, $q=2$, and $q=3$ models retained 100,000, 150,000, and 200,000 MCMC samples, respectively. MCMC convergence was assessed by inspecting the marginal posterior distributions and verifying their stability across early, middle, and late post-burn-in samples.

\section{Results and Discussions}
The BIC values for the $q=1$, $q=2$, and $q=3$ models were -7018, -7030, and -7011, respectively. The $q=2$ model, comprising two finite sediment layers over a half-space, therefore provides the preferred balance between data fit and model complexity. The additional parameters of the $q=3$ model are not supported by the RDC measurements.

\begin{figure}[htbp]
    \centering
    \includegraphics[width=0.7\linewidth]{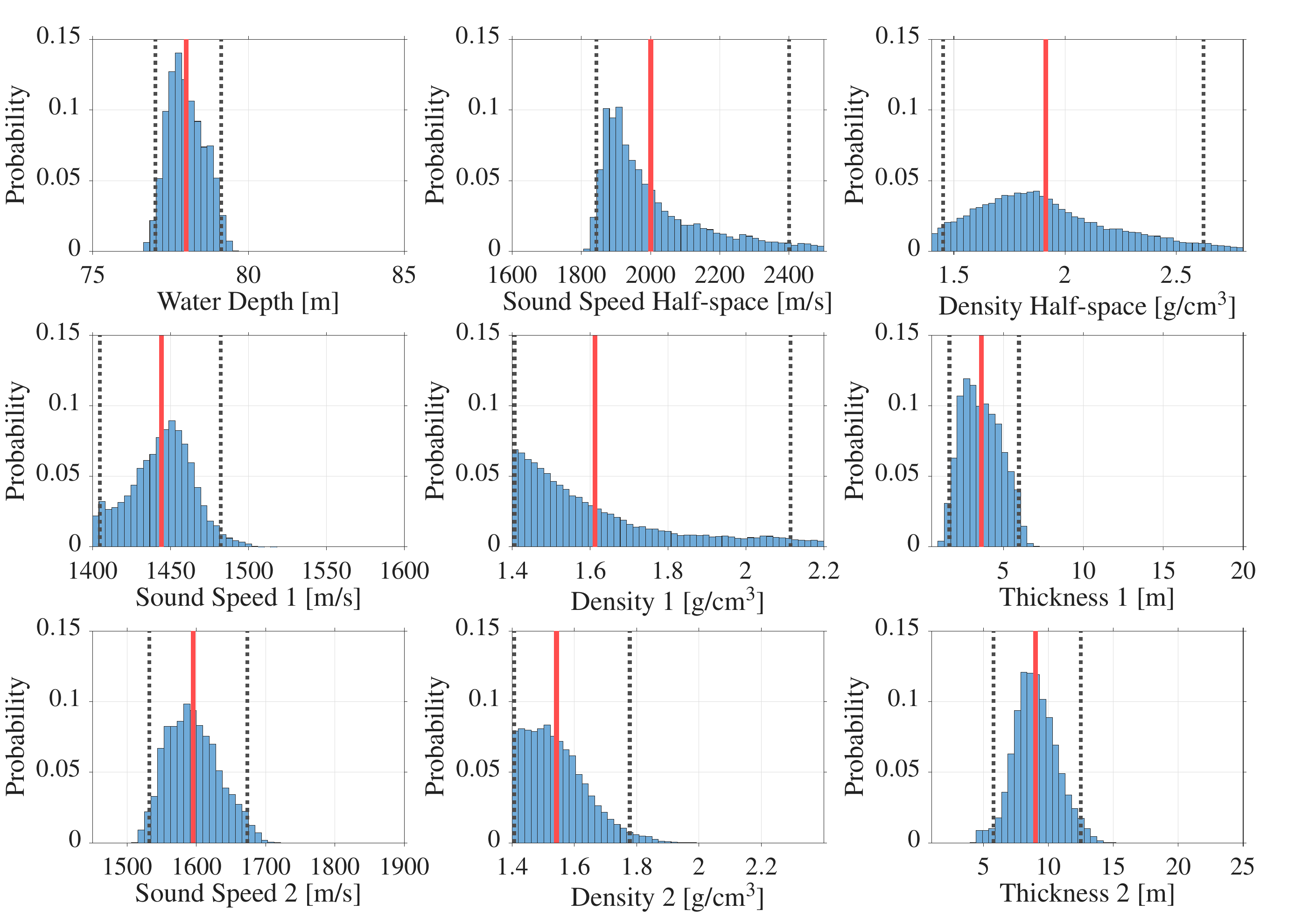}\hfill % \vspace{-.5in}
    \caption{Marginal posterior distributions for the preferred $q=2$ geoacoustic model, comprising two finite sediment layers over a half-space. The red solid lines denote the MMSE estimate, and the black dotted lines indicate the 95\% credible intervals. The horizontal axis of each panel spans the prior bounds of the corresponding parameter. The ordinate represents probability mass per histogram bin.} % \vspace{-.1in}
    \label{fig:post_results}
    
\end{figure}

Fig.~\ref{fig:post_results} shows the marginal posterior distributions, and Table~\ref{tab:results} summarizes the prior bounds and inversion results for the preferred $q=2$ model. The first- and second-layer sound speeds and thicknesses are relatively well constrained, whereas the half-space parameters exhibit broader posteriors, indicating decreasing sensitivity with depth. The inferred effective water depth is 78.0~m, consistent with the approximately 77--79~m bathymetric variation along the analyzed track. Sediment densities are weakly constrained, consistent with RDC locations being primarily sensitive to modal dispersion rather than amplitudes. Their broad, skewed posterior distributions, with 95\% CIs spanning much of the prior range, warrant caution in interpreting the apparent depth-dependent variability in the MMSE density profile. The predicted RDCs agree well with the observations (Fig.~\ref{fig:data_ex}(c)). Residuals are generally consistent with the assumed Gaussian error model, with only modest nearest-neighbor correlation. Residual histograms and sample autocorrelation functions for each mode pair are provided in Fig.~S1 of the Supplementary Material.

\begin{figure}[htbp]
    \centering
    \includegraphics[width=0.65\linewidth]{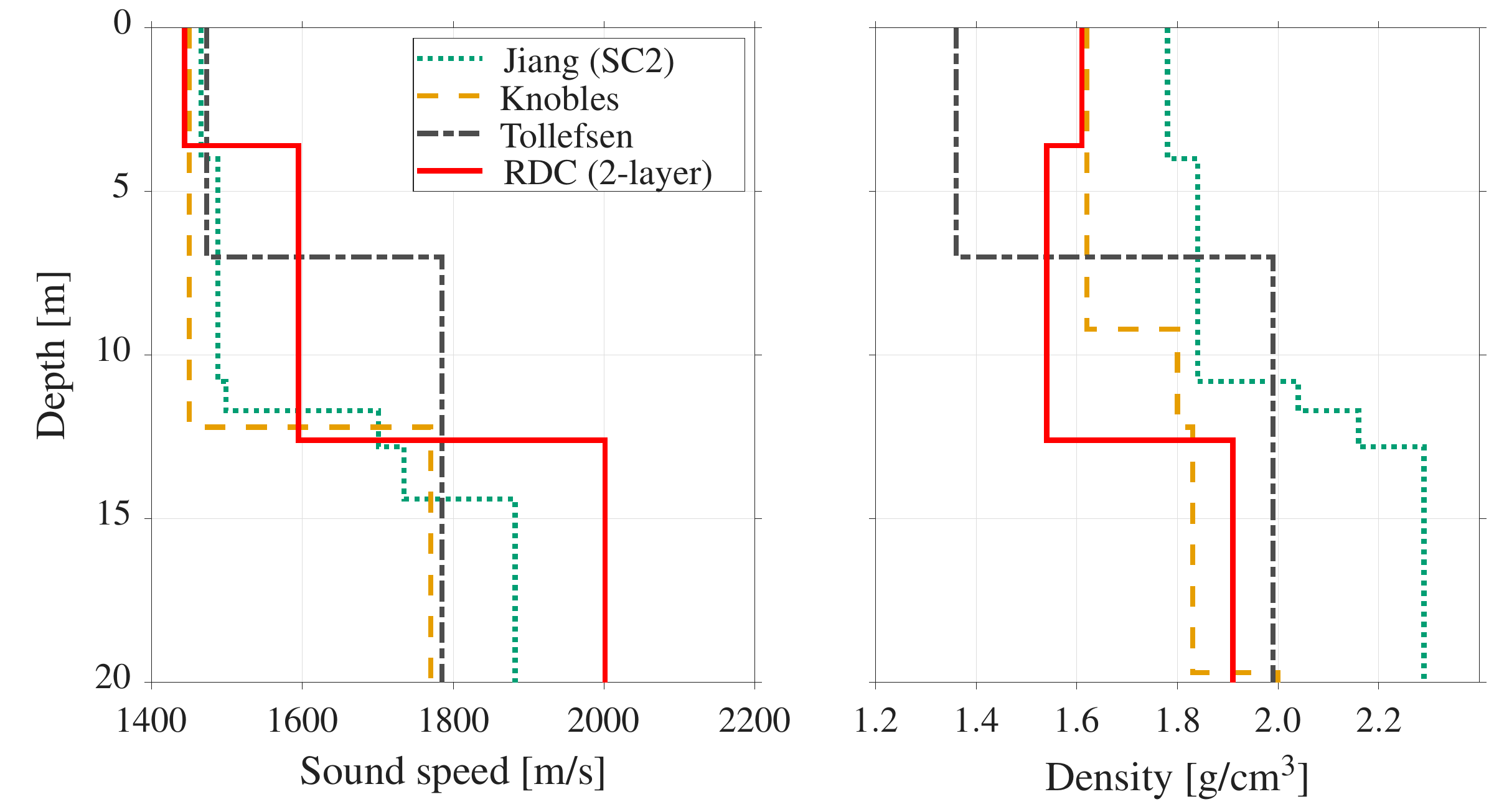}\hfill
    \caption{Comparison of representative geoacoustic profiles from the SC2 reflection-coefficient inversion of Jiang et al.\cite{JiaHolDosDet:J23} (green dotted), Knobles et al.\cite{KnoNeiWilHodBonLin:J22} (orange dashed), Tollefsen et al.\cite{TolHodDosBonKno:J22} (gray dash-dot), and the BIC-preferred ($q=2$) RDC inversion from this work (red solid), comprising two finite sediment layers over a half-space.}
    \label{fig:results_comp}
\end{figure}

To assess the physical consistency of the inferred sediment structure, Fig.~\ref{fig:results_comp} compares the present result with the reflection-coefficient inversion of Jiang et al.~\cite{JiaHolDosDet:J23} at site SC2, located within a few hundred meters of VLA2, and the ship-noise inversions of Knobles et al.~\cite{KnoNeiWilHodBonLin:J22} and Tollefsen et al.~\cite{TolHodDosBonKno:J22}. The latter two studies are particularly relevant because they analyzed ship noise recorded on the same VLA during SBCEX17. Knobles et al. inferred geoacoustic properties from waveguide-invariant singular-point features extracted from single-hydrophone ship spectrograms, whereas Tollefsen et al. performed a Bayesian matched-field inversion of the complex acoustic field using all 16 elements of the VLA2 and a discrete set of narrowband and broadband frequency components extracted from an 8-s data segment centered at the CPA.

% Tab: MCMC search space
\begin{table}[htbp]
\caption{\label{tab:results} Minimum mean-square error (MMSE) estimates and 95\% credible intervals (CIs) for the preferred $q=2$ geoacoustic model.}

\begin{ruledtabular}
\begin{tabular*}{\linewidth}{@{\extracolsep{\fill}}l ccc ccc ccc}
% \begin{tabular}{l ccc ccc ccc}
\multirow{2}{*}{Parameter}
& \multicolumn{3}{c}{Depth/thickness [m]}
& \multicolumn{3}{c}{Sound speed [m/s]}
& \multicolumn{3}{c}{Density [g/cm$^{3}$]} \\
\cline{2-4}\cline{5-7}\cline{8-10}
& Mean & CI & Prior bounds & Mean & CI & Prior bounds & Mean & CI & Prior bounds\\
\hline
Water
& 78.0 & [77.1, 79.1] & [75.0, 85.0]
& -- & -- & -- 
& -- & -- & -- \\

Layer 1
& 3.6 & [1.66, 5.93] & [0.5, 20.0]
& 1444 & [1403, 1481] & [1400, 1600]
& 1.61 & [1.41, 2.07] & [1.40, 2.20]\\

Layer 2
& 9.0 & [6.2, 12.5] & [1.0, 25.0]
& 1595 & [1536, 1671] & [1450, 1900] 
& 1.54 & [1.41, 1.74] & [1.40, 2.40] \\

Half-space
& -- & -- & --
& 2001 & [1844, 2422] & [1600, 2500] 
& 1.91 & [1.45, 2.56] & [1.40, 2.80] \\
\end{tabular*}
\end{ruledtabular}
\end{table}

%  1) surficial sound speed value in the upper mud (you do really good)
The preferred $q=2$ inversion yields a surficial layer with a sound speed of 1444~m/s, in good agreement with the surficial mud sound speed inferred by the active reflection-coefficient inversion of Jiang et al.~\cite{JiaHolDosDet:J23}. This agreement is the most direct validation of the present inversion because it is the parameter most directly comparable with the local SC2 result. The inferred surficial-layer thickness of 3.6 m is smaller than the mud thickness local at SC2\cite{JiaHolDosDet:J23} and the interface depth of approximately 7 m inferred by Tollefsen et al.\cite{TolHodDosBonKno:J22} near CPA (Fig.~\ref{fig:results_comp}). This is expected because the present inversion uses RDCs extracted over a several-kilometer propagation path, along which the mud thins with increasing range from VLA2 (Fig.~\ref{fig:scenario}). The inferred thickness represents a path-averaged effective value.

Because their singular points are sparse, Knobles et al.\cite{KnoNeiWilHodBonLin:J22} assigned fixed thicknesses (9.2 and 3 m) to the mud and transition layers but represented them by a common low-sound-speed parameter, effectively a single low-speed unit; similarly, Tollefsen et al.\cite{TolHodDosBonKno:J22} inferred a single dominant interface at approximately 7 m depth using the full array. In contrast, the RDCs constrain the surficial-layer thickness directly and support an additional effective layer using a single hydrophone. This second layer (1595 m/s) represents an effective transition from the low-speed surficial mud toward higher-speed sediments rather than a distinct geological unit, and its base at 12.6 m is consistent with the combined 12.2 m mud and transition thickness assumed by Knobles et al.

\section{Conclusions}
We demonstrated passive geoacoustic inversion using RDCs extracted from continuous ship noise recorded on a single hydrophone. Bayesian inversion combined with the BIC selected a model with two finite sediment layers over a half-space; the inferred surficial-mud sound speed agreed well with a local active-source inversion, and the RDCs supported an additional effective layer relative to previous ship-noise inversions. These results indicate that modal wavenumber differences, together with the required auxiliary environmental and source information, contain sufficient information to recover the dominant seabed structure in shallow water with substantially reduced sensing requirements.

Several limitations remain. First, the present study assumes that the mode-pair identities of the observed RDCs are known, which isolates the information content of the RDCs from the identification problem. An incorrect assignment would bias the inversion, and automated mode-pair identification, for example through optimization\cite{BonnelEtal:J18} and probabilistic data association\cite{MeyKroWilLauHlaBraWin:J18}, is required for operation without prior environmental information. Second, the forward model assumes a range-independent waveguide, whereas the mud thins along the propagation path. The inferred parameters therefore represent path-averaged effective values. For example, our estimate of the surficial-layer thickness (3.6 m) represents the area between the receiver and the ship track, but it differs from the local mud thickness ($\sim\rmv$12 m) at the receiver position. Finally, because the inversion relies solely on RDC locations, the sensitivity to sediment density and attenuation is limited.

\section*{Supplementary Material}
See supplementary material at [URL will be inserted by AIP] for residual diagnostics and evaluation of the measurement-error model.

\begin{acknowledgments}
We thank Dr. W. S. Hodgkiss for providing the dataset. This work was supported by a Woods Hole Oceanographic Institution Postdoctoral Scholarship and Office of Naval Research Grants N00014-24-1-2021 and N00014-26-1-2087.
\end{acknowledgments}

\section*{Author Declaration}
The authors have no conflicts to disclose.

\section*{Data Availability}\label{sec:DataAvailability}
The acoustic data that support the findings of this study were collected by the \textit{Marine Physical Laboratory}. Data are available from the authors upon reasonable request.

\bibliography{IEEEabrv,StringDefinitions,SALBooks,SALPapers,Temp_V2}

\end{document}